\documentclass[aps,amsmath,amssymb,prapplied,reprint,onecolumn,superscriptaddress,nofootinbib]{revtex4-2}

\usepackage{graphicx}
\usepackage{empheq}
\usepackage{tikz}
\usepackage{bm}
\usepackage{bbm}
\usepackage[english]{babel}
\usepackage[utf8]{inputenc}
\usepackage[top=1in, bottom=1in, left=1.25in, right=1.25in]{geometry}

\usepackage{empheq}
\usepackage{hyperref}

\newcommand{\ket}[1]{|#1\rangle}
\newcommand{\bra}[1]{\langle#1|}

\newcommand{\sz}[0]{\sigma^z}
\newcommand{\sx}[0]{\sigma^x}

\newcommand{\eps}[0]{\epsilon}

\begin{document}
\title{Preparation of a Bell state in an analogue device hosting the transverse-field Ising model}
\author{Ana Palacios}
\email{ana.palacios@qilimanjaro.tech}
\affiliation{%
 Qilimanjaro Quantum Tech, Carrer de Veneçuela, 74, Sant Martí, 08019, Barcelona, Spain}
\affiliation{Departament de F\'{i}sica Qu\`{a}ntica i Astrof\'{i}sica, Facultat de F\'{i}sica,
Universitat de Barcelona, 08028 Barcelona, Spain}
\affiliation{Institut de Ci\`{e}ncies del Cosmos, Universitat de Barcelona,
ICCUB, Mart\'{i} i Franqu\`{e}s 1, 08028 Barcelona, Spain.}

\date{\today}

\begin{abstract}
    We propose a protocol for preparing the singlet Bell state in an analogue device implementing the transverse-field Ising model with ZZ interactions and local X and Z fields, which can be engineered within the superconducting platform, among others. This protocol is performed by direct control of the terms of the Hamiltonian alone. The method exploits the fact that the singlet state is the first excited state of a symmetric Hamiltonian family, and comprises two steps: an adiabatic interpolation preparing the ground state of said symmetric Hamiltonian, followed by the resonant population transfer between its two lowest energy levels. For realistic parameters, the protocol achieves fidelities comparable to standard gate-based preparation within similar time scales ($\sim$100 ns), and can reach infidelities on the order of $10^{-4}$ with a moderately increased duration. We further analyse robustness against systematic control errors and show that high fidelities are maintained under implementation imperfections.
\end{abstract}

\maketitle

\section{Introduction}

Entanglement lies at the heart of quantum information science, enabling protocols with no classical analogue such as quantum teleportation, superdense coding, entanglement-based quantum key distribution, and device-independent cryptography~\cite{bennett_teleporting_1993,bennett_communication_1992,ekert_quantum_1991,horodecki_quantum_2009}. Among entangled states, two-qubit Bell states constitute the simplest and most fundamental resource. They serve both as operational primitives in quantum communication and as benchmark states for calibrating and validating quantum processors. Consequently, the rapid and high-fidelity preparation of Bell states is a central requirement across quantum technologies.

In most hardware platforms, Bell-state preparation is implemented within the gate-based paradigm of quantum computation, typically via a sequence of single-qubit rotations and an entangling two-qubit gate (e.g., CNOT or CZ). In contrast, analogue approaches to entangled-pair generation are more common in photonic systems, where spontaneous parametric down-conversion provides a direct mechanism for Bell-state generation without digital control~\cite{kwiat_new_1995}. While such optical methods can suffer from low efficiency, they illustrate that entanglement generation need not be intrinsically gate-based.

More broadly, analogue and digital–analogue quantum devices are currently at the forefront of experimental quantum science, producing some of the most interesting recent experiments in the superconducting platform~\cite{king_quantum_2023,king_beyond-classical_2025,hu_overcoming_2024,karamlou_probing_2024,rosen_flat-band_2025}.
All these recent experiments have demonstrated the power of Hamiltonian engineering as an alternative to strictly gate-based computation. In this context, restricting Bell-state preparation to digital circuits may represent an unnecessary limitation. Analogue generation of entangled pairs, be it within a hybrid digital–analogue architecture or in a dedicated ``entanglement factory'' chip distributing Bell pairs to a processor, offers an attractive and largely unexplored alternative.

Some analogue entanglement-generation protocols have been proposed in specific platforms, such as Rydberg atoms exploiting blockade mechanisms~\cite{wilk_entanglement_2010}. However, these schemes are often tightly linked to the microscopic level structure of the system and do not readily generalize to other architectures.

Here we present a protocol for preparing Bell states using purely analogue control in a device described by the transverse-field Ising model with ZZ interactions and local X and Z fields, which are the control primitives commonly available in superconducting architectures. The protocol operates entirely within the computational subspace and is therefore, in principle, transferable to any qubit-emulation platform implementing similar interactions.

Since our focus is on superconducting implementations, we benchmark the performance of the protocol against standard gate-based preparation. Contemporary superconducting processors routinely achieve single-qubit (1QB) gate fidelities above 0.999 with durations in the 10–30 ns range.
The more intricate step are two-qubit (2QB) gates, which are ones introducing most of the error. While ultra-fast 2QB gates (like~\cite{xu_high-fidelity_2020} with 30ns and fidelity of 0.995 and~\cite{an_zz-free_2025} with 20ns and fidelity of 0.996) or ultra-high fidelity 2QB gates (like~\cite{li_realization_2024} with 48ns and fidelity of 0.9990) have been demonstrated in optimised devices, standard implementations are typically around 70-100 ns. For example, as of the time of writing ibm\_boston has 32ns 1QB gates and 68ns 2QB gates, with median fidelities of 0.9998 and 0.9985, respectively. A digital preparation of a Bell state thus typically requires $\sim$80–130 ns and results in errors on the order of $10^{-3}$.

We show that our analogue protocol achieves comparable preparation times and equal or improved fidelities under realistic parameters, while offering a conceptually distinct route to entanglement generation based purely on Hamiltonian control.

\section{Preparation of the singlet state}

\begin{figure}[ht]
    \centering
    \includegraphics[width=0.4\linewidth]{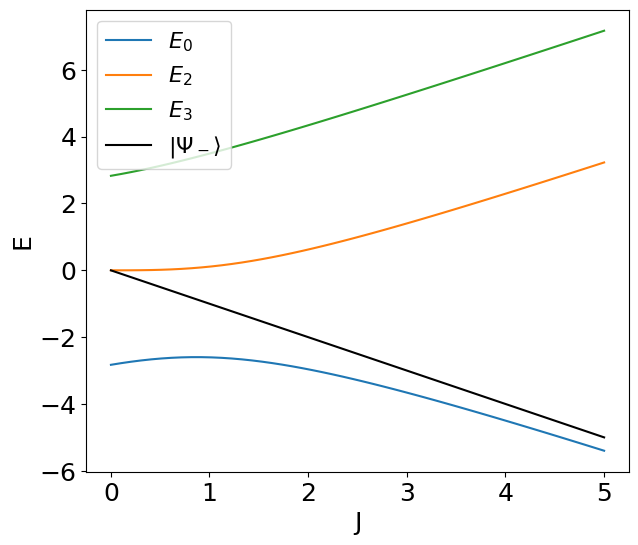}
    \caption{Energy spectrum of $H$ as a function of the coupling strength $J$.}
    \label{fig:spectrum_H_vsJ}
\end{figure}

Let us consider the particularly symmetric Hamiltonian
\begin{equation}
H = (\sigma^x_0 + \sigma^x_1 + \sigma^z_0 + \sigma^z_1) + J \sigma^z_0 \sigma^z_1 
\end{equation}
with $J > 0$. The method proposed here stems from the observation that the first excited state of said Hamiltonian is precisely the singlet state, 
\begin{equation}
    \ket{\Psi_-} = \frac{\ket{01} - \ket{10}}{\sqrt{2}} \ ,
\end{equation}
with $H\ket{\Psi_-} = -J \ket{\Psi_-}$. This is true independently of the value of the coupling strength $J>0$; for $J<0$, $\ket{\Psi_-}$ is the second excited state instead of the first one. The rest of the spectrum can also be obtained as an analytical function of $J$ from the roots of the characteristic polynomial of $H$ when projected onto $P_{\text{rest}} = \mathbbm{1} - \ket{\Psi_-}\bra{\Psi_-}$, which is 
\begin{equation}
    -\lambda^3 + J\lambda^2 + (J^2 + 8)\lambda - J^3 = 0 \ .
    \label{eq:characteristic_poly}
\end{equation}
The shape of the spectrum over a range of $J$ is depicted in Fig.~\ref{fig:spectrum_H_vsJ}.

Thus, our aim will be to prepare this first excited state. We will assume the procedure is initialised with perfect fidelity in the ground state of Hamiltonian $H_0 = \sigma_0^x + \sigma_1^x$ (with ground state $\ket{\varphi_0} = \ket{--}$), which is a standard initial Hamiltonian in many analogue superconducting processes, such as those related to analogue quantum optimisation. 

\begin{figure}[ht]
\begin{tikzpicture}
      \node[anchor=south west, inner sep=0] (img) at (0,0)
        {\includegraphics[width=0.45\linewidth]{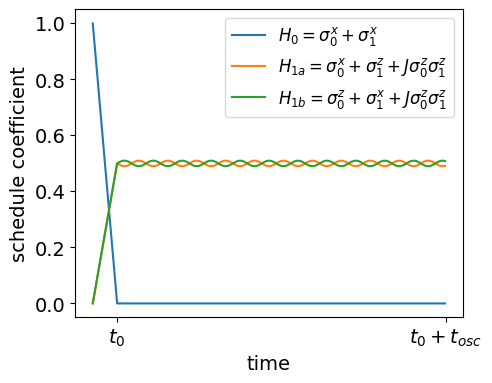}
        \includegraphics[width=0.43\linewidth]{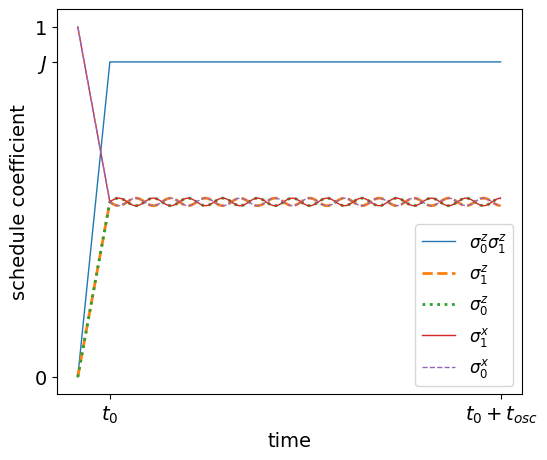}};
      \begin{scope}[x={(img.south east)}, y={(img.north west)}]
        \node[] at (0.15,0.82) {a)};
        \node[] at (0.93,0.75) {b)};
      \end{scope}
    \end{tikzpicture}
    \caption{Representation of the schedule for the whole process of Bell state preparation. In a) we present the schedule from an algorithmic perspective (participating Hamiltonians), while in b) we show the schedule in terms of controls of the different terms of the Hamiltonians, more relevant to the hardware implementation.}
    \label{fig:schedule_agnostic}
\end{figure}

The summary of the protocol is as follows: we first prepare the ground state of $H$ via adiabatic interpolation for time $t_0$, described in Section~\ref{sec:adiabatic_interp}. Then, we carry out a population transfer between ground and first excited states by resonantly driving said transition via an oscillating schedule for a time $t_{osc}$, the details of which are presented in Section~\ref{sec:rabi_osc}. The resulting schedule is illustrated in Fig.~\ref{fig:schedule_agnostic}.

\subsection{Step 1: adiabatic interpolation}\label{sec:adiabatic_interp}

The first thing to note is that $P_{\Psi_-} = \ket{\Psi_-} \bra{\Psi_-}$ commutes with $H_0$; thus, no direct interpolation between $H_0$ and $H$ will populate $\ket{\Psi_-}$, as its presence or absence is a conserved quantity of motion and it has zero overlap with $\ket{\varphi_0}$. Consequently, the energy difference relevant to the adiabatic condition is in this case $\Delta_{02} = E_2 - E_0$, where $E_i$ are the instantaneous eigenvalues of $H_{\text{interp}}$,
\begin{gather}
    H_{\text{interp}}(s) = (1-s) H_0 + s H_1 \\
    H_1 = \frac{1}{2} H(2J)
\end{gather}
for $s=t/t_0$ and $t_0$ the duration of the interpolation. The particular relation of $H_1$ to $H$ arises in the interest of convenience for the description of the second part of the protocol.

\begin{figure}[ht]
\begin{tikzpicture}
      \node[anchor=south west, inner sep=0] (img) at (0,0)
        {\includegraphics[width=0.8\linewidth]{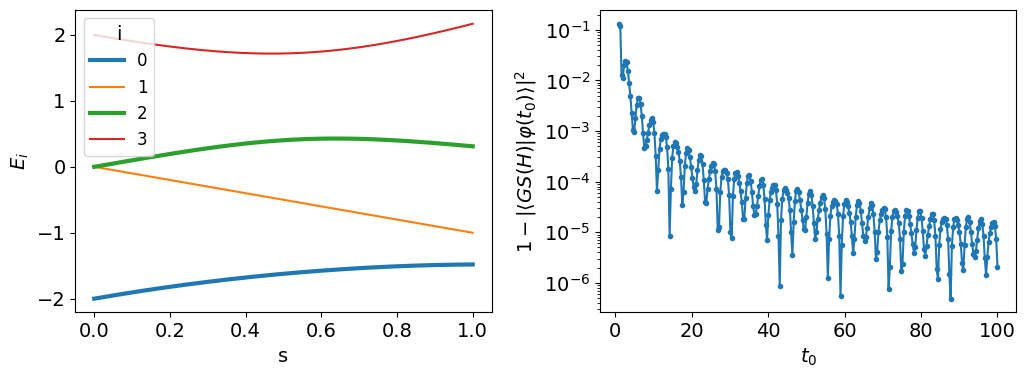}};
      \begin{scope}[x={(img.south east)}, y={(img.north west)}]
        \node[] at (0.42,0.8) {a)};
        \node[] at (0.93,0.8) {b)};
      \end{scope}
    \end{tikzpicture}
    \caption{Energy landscape of $H_{\text{interp}}$ (a) and final infidelities at $s=1$ as a function of interpolation time $t_0$ (b) for the first step of the protocol, considering $J=1$. In a), levels 0 and 2 are thickened to highlight the relevant gap for the adiabatic process, which is minimum at $s=1$ ($E_2(1) - E_0(1) \simeq 1.8$).}
    \label{fig:H_interp_spectrum}
\end{figure}

The energy spectrum of $H_{\text{interp}}$ is illustrated in Fig.~\ref{fig:H_interp_spectrum}.a for a fixed $J$. The resulting infidelity with the ground state of $H$ for different interpolation times $t_0$ is shown in Fig.~\ref{fig:H_interp_spectrum}.b, where it becomes apparent that from $t_0\simeq 10$ onwards it is possible to obtain errors below $10^{-4}$ for the illustrated example ($J=1$). The fidelity with which we prepare this ground state sets an upper limit to the fidelity with which we can reach the Bell state at the end of the protocol.

Since we can project onto the subspace given by $P_{\text{rest}}=\mathbbm{1} - \ket{\Psi_-}\bra{\Psi_-}$, one can even obtain an analytical description of the gap as a function of $s$ (Eq.~\eqref{eq:characteristic_poly}) and use this information to optimise the adiabatic time $t_0$ beyond that required by a simple linear interpolation. However, in the interest of simplicity we limit ourselves to a linear schedule in this analysis.

\subsection{Step 2: population transfer}\label{sec:rabi_osc}

Once we are in the ground state of our target Hamiltonian $H$, all that remains is going up one step in the ladder of eigenstates. If we consider the two lowest levels to be sufficiently isolated from the rest (and, in addition, that their energy difference $\Delta_{01} = E_1 - E_0$ is sufficiently far from being an exact divisor of other transitions involving $E_1, E_0$), this becomes a standard population inversion problem in a two-level system scenario \cite{sakurai_modern_2011}. Consequently, we would like to induce half of a Rabi oscillation. However, in our analogue device we do not have a resonator coupled to the eigenstates of the two-level system of interest, as we would have in a standard single-qubit gate implementation, for example; here our controls are exclusively on the terms of the Hamiltonian and, furthermore, the two levels of interest are many-body states. 
The observation that becomes central to this task is that there is a convenient partition of $H$ such that, if independently controlled, can enable the transition between $E_0$ and $E_1$ in a similar way the coupling to a resonator does for a digital gate in a superconducting qubit. Said partition is the following:
\begin{gather}
    H(2J) = H_{1a}(J) + H_{1b}(J) \\
    H_{1a}(J) = \sx_0 + \sz_1 + J\sz_0 \sz_1 \\
    H_{1b}(J) = \sz_0 + \sx_1 + J\sz_0 \sz_1
\end{gather}

\begin{figure}[ht]
\begin{tikzpicture}
      \node[anchor=south west, inner sep=0] (img) at (0,0)
        {\includegraphics[width=0.8\linewidth]{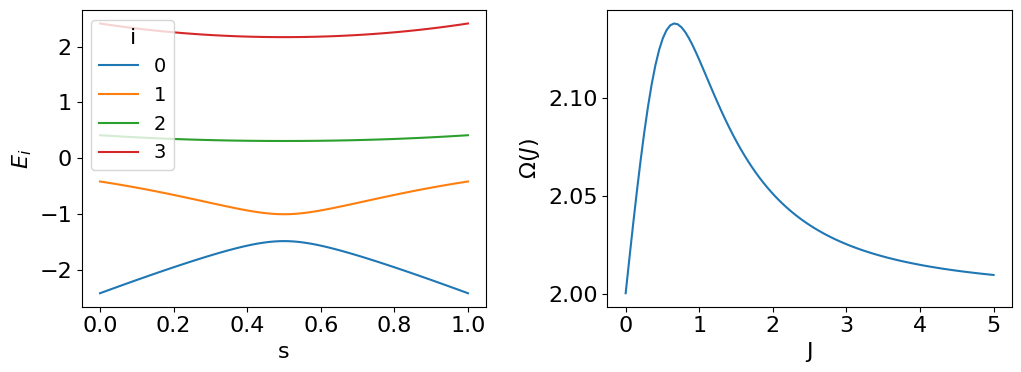}};
      \begin{scope}[x={(img.south east)}, y={(img.north west)}]
        \node[] at (0.42,0.8) {a)};
        \node[] at (0.93,0.8) {b)};
      \end{scope}
    \end{tikzpicture}
    \caption{Energy landscape of $H_2$ for $J=1$ (a) and resulting transition matrix element between the first two lowest eigenstates of $H_1$ induced by the oscillation around $s^*=0.5$ (i.e., around $H_1$) (b).}
    \label{fig:specH2_J1__transition_el_vs_J}
\end{figure}

Now, if we consider the interpolation between $H_{1a}, H_{1b}$,
\begin{equation}
    H_2(s) = (1-s)H_{1a} + s H_{1b}
    \label{eq:H2_interp}
\end{equation}
we can observe that the minimum gap point in this interpolation is at $s^* = 0.5$ (see Fig.~\ref{fig:specH2_J1__transition_el_vs_J}a), where $H_2(s^*=0.5) = \frac{1}{2}H(2J) = H_1(J)$. This is useful to our purposes because it means that, to first order, the energy separation between ground and first excited states doesn't change when perturbing around $s^*$. The fact that this holds for any value of $J$ can be checked via a perturbation theory analysis around $s^*$, considering the unperturbed Hamiltonian to be $H_1$ and $V = H_{1b} - H_{1a} = (\sz_0 - \sx_0) - (\sz_1 - \sx_1)$ as the perturbation. The description of $V$ in the Bell basis amounts to
\begin{equation}
    V = 2 \left[\left(\ket{\Psi_+} \bra{\Psi_-} + \ket{\Phi_-} \bra{\Psi_-}\right) + \text{h.c.}\right]
    \label{eq:pert_in_Bell_basis}
\end{equation}
Since $\ket{E_0^{(0)}} = \alpha \ket{\Psi_+} + \beta\ket{\Phi_-} + \gamma \ket{\Phi_+}$ for some $\alpha, \beta, \gamma \in \mathbb{C}$ and $\ket{E_1^{(0)}} = \ket{\Psi_-}$, it is easy to see that $\bra{E_1^{(0)}} V \ket{E_1^{(0)}} = \bra{E_0^{(0)}} V \ket{E_0^{(0)}} = \Delta_{01}^{(1)} =  0$.
Thus, we can implement a full population inversion via an oscillating schedule around $s^*$ in a way that is completely analogous to a Rabi pulse. 
\begin{gather}
H_{2, osc} (t) = H_1 +  \varepsilon \sin{\omega t} \ (H_{1b} - H_{1a})
\end{gather}
where we have set $s= s^* + \varepsilon \sin{\omega t}$ in~\eqref{eq:H2_interp} and with $0 < \varepsilon \ll \Delta_{01}$, such that our perturbative description to first order is correct and the gap remains constant~\footnote{Note that this is analogous to standing in the dispersive regime in typical gate-based schemes.}.
When we project onto the two lowest levels, the effective Hamiltonian of the evolution becomes:
\begin{gather}
    h^*(t) = \frac{\Delta_{01}}{2} \sigma^z + \varepsilon \Omega \sin{(\omega t)}  \   \sx 
    \label{eq:effective_2level_ham}\\
    \Omega = |\bra{E_1^{(0)}} V \ket{E_0^{(0)}}|
\end{gather}
From Eq.~\eqref{eq:pert_in_Bell_basis} it can be readily seen that the transition matrix element between $\ket{E_0^{(0)}}$ and $\ket{E_1^{(0)}}$ is
\begin{equation}
    \bra{E_1^{(0)}} V \ket{E_0^{(0)}} = 2 (\alpha + \beta)
\end{equation}
The dependence of $|\bra{E_1^{(0)}} V \ket{E_0^{(0)}}|$ on $J$ is shown in Fig.~\ref{fig:specH2_J1__transition_el_vs_J}b, where it can be seen that $\Omega \simeq 2$ across all $J$.

From Eq.~\eqref{eq:effective_2level_ham} we can follow the standard Rabi theory to find the duration of the oscillation that gives us a full population inversion, $t_{osc} = \frac{\pi}{\Omega \varepsilon}$, and drive frequency, which should be at resonance, $\omega = \Delta_{01}$. 

\begin{figure}[ht]
\begin{tikzpicture}
      \node[anchor=south west, inner sep=0] (img) at (0,0)
        {\includegraphics[width=0.45\linewidth]{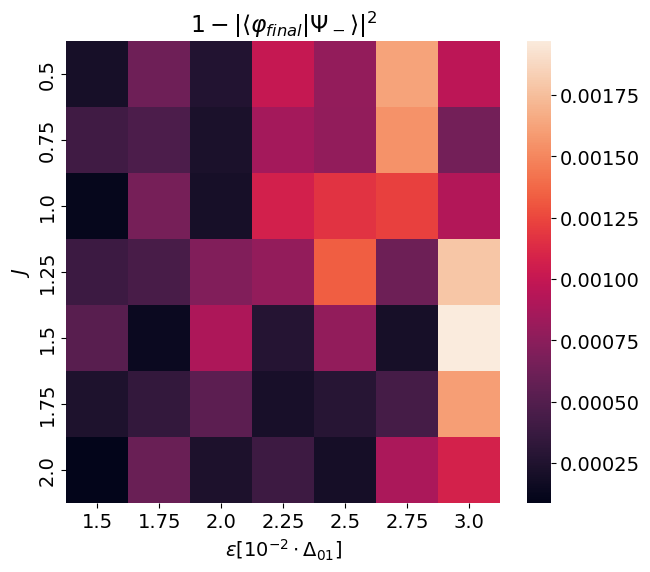}
        \includegraphics[width=0.45\linewidth]{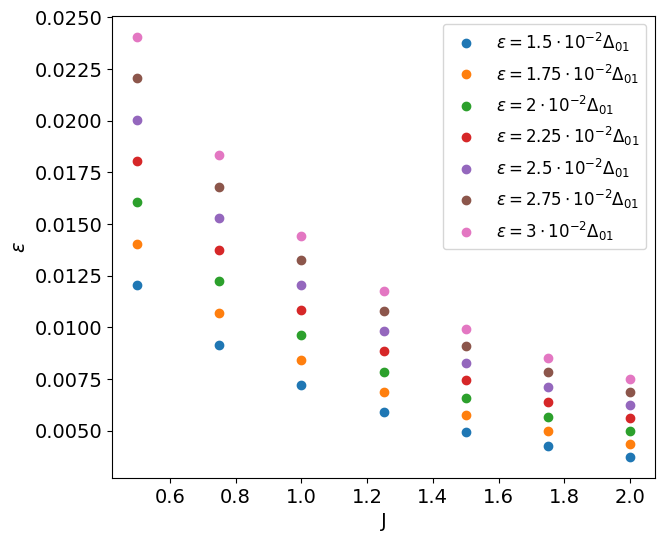}};
      \begin{scope}[x={(img.south east)}, y={(img.north west)}]
        \node[] at (0.05,0.97) {a)};
        \node[] at (0.57,0.97) {b)};
      \end{scope}
    \end{tikzpicture}
    \caption{In a) we show the final infidelity with the target singlet state over a range of parameters. For each $J$, the optimal $t_0$ is selected over a grid of 10 equally spaced points between 10 and 20 to account (to some degree) for the oscillations in the ground state fidelity of the adiabatic preparation (see Fig.~\ref{fig:H_interp_spectrum}.b), but in every case $t_0\simeq 20$. Figure b) shows the absolute value of the corresponding amplitudes $\varepsilon$ as a function of $J$.}
    \label{fig:final_fid_parameter_range}
\end{figure}

\begin{figure}[!ht]
    \centering
    \includegraphics[width=0.6\linewidth]{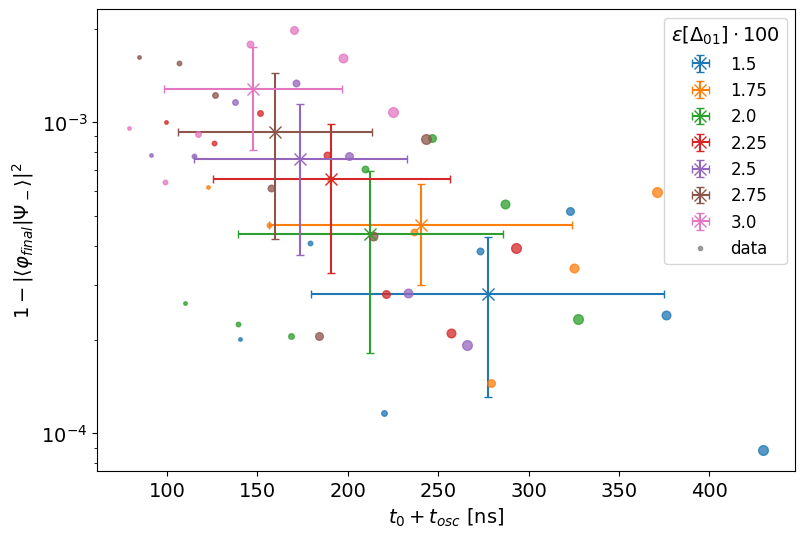}
    \caption{Achieved infidelity versus the total time, considering the energy to be in GHz units. Circles correspond to the simulated data, with larger sizes indicating larger $J$ values and their colour referring to their $\varepsilon$ value. The crosses and associated error bars indicate the mean and standard deviation over all $J$ values for a given $\varepsilon$. These are intended as a guide for the eye to highlight the tendency of the data.}
    \label{fig:infid_vs_totaltime_units_pretty}
\end{figure}

The ideal implementation of the full protocol for $J=1$, interpolation time $t_0=20$ and oscillation amplitude $\varepsilon=0.02\cdot \Delta$ provides a final fidelity of $|\bra{\varphi_{final}}\Psi_-\rangle|^2 = 0.9998$. 
If we consider the coupling to be in GHz, the corresponding units of time are nanoseconds, which amounts to a competitive performance versus standard digital preparation methods. As stated in the introduction, in a generic reasonably high-quality setting a Bell state can be generated up to 0.1\% error in around 100ns; with our protocol, we can achieve roughly an order of magnitude improvement with only an increase of a factor of two in the time budget. To strengthen this claim and show that it is not limited to the choice of $J=1$GHz, we study the final fidelity with the target state over a broader range of parameters in Fig.~\ref{fig:final_fid_parameter_range}.a. The latter shows that smaller amplitudes provide higher accuracies but also longer preparation times, as $t_{osc}$ is inversely proportional to $\varepsilon$. This effect can be readily observed in Fig.~\ref{fig:infid_vs_totaltime_units_pretty}, where we have set the energy scale to GHz and thus fixed the absolute protocol times to ns units. 

The interplay between accuracy and speed of the protocol is mainly controlled by the oscillation amplitude $\varepsilon$; smaller amplitudes lead to higher accuracy but longer state preparation times. This is, indeed, apparent in Fig.~\ref{fig:final_fid_parameter_range}.a, and made more explicit in the tendencies presented in Fig.~\ref{fig:infid_vs_totaltime_units_pretty}, where we average with respect to $J$. At the same time, the coupling strength $J$ determines when $\varepsilon$ is sufficiently small for the resonance condition to be respected, thus setting the upper bound to $\varepsilon$ for a successful population transfer (see Fig~\ref{fig:final_fid_parameter_range}.b). Furthermore, for $J$ too large, the gap at $s^*$, $\Delta_{01}$, may become so small (see Fig.~\ref{fig:spectrum_H_vsJ}) that the precision required to oscillate with $\varepsilon$ (and the very frequency $\Delta_{01}$) may exceed the capabilities of the setup. Finally, the minimum adiabatic interpolation time to achieve a certain fidelity with the ground state of $H_1$ also depends on $J$, but the main contribution to the time budget comes from the oscillation time $t_{osc}$.

\section{Robustness to non-idealities}

In an experimental implementation, there's a broader range of error sources that need to be taken into account. Since standard coherence times in superconducting qubits are around tens of microseconds, about two orders of magnitude above the time our protocol takes (which we realistically consider to be of the order of 100 ns), we won't consider environmental noise as the main source of error, since the device should remain fairly coherent along the preparation of the target state. Nonetheless, since we are interested in studying high accuracies, its effect should be studied in a more complete characterisation, which we leave for future work. Instead, since our protocol relies heavily on the symmetry between the X and Z local fields, which is essential for $\ket{\Psi_-}$ to be an eigenstate of $H_1$, we focus the analysis on the robustness to control errors. These control errors will prevent us from guaranteeing the perfect balance of all local fields, which will in turn affect the fidelity with which we can prepare our singlet states. 

We study the impact of this error source by considering the presence of an unknown bias on each term of the Hamiltonian, such that we have the following model:
\begin{gather}
    H_0^\prime = (1+\eta\eps_0) \sx_0 + (1+\eta\eps_1) \sx_1 \\
    H_{1a}^\prime = (1+\eta\eps_0) \sx_0 + (1+\eta\eps_2) \sz_1 + J(1 + \eta\eps_4)\sz_0 \sz_1 \\
    H_{1b}^\prime = (1+\eta\eps_3) \sz_0 + (1+\eta\eps_1) \sx_1 + J(1 + \eta\eps_4)\sz_0 \sz_1 
\end{gather}
where the $\eps_i\in [-0.5, 0.5]$ are drawn from a random uniform distribution and $\eta$ characterises the noise strength.

\begin{figure}[ht]
\begin{tikzpicture}
      \node[anchor=south west, inner sep=0] (img) at (0,0)
        {\includegraphics[width=0.49\linewidth]{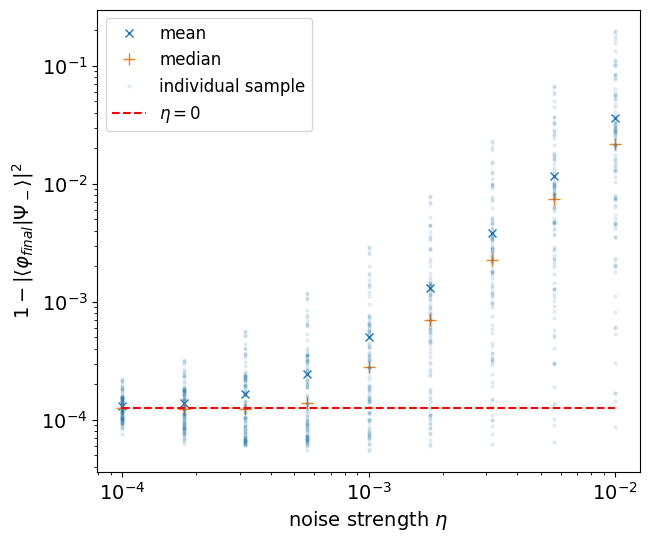}
        \includegraphics[width=0.49\linewidth]{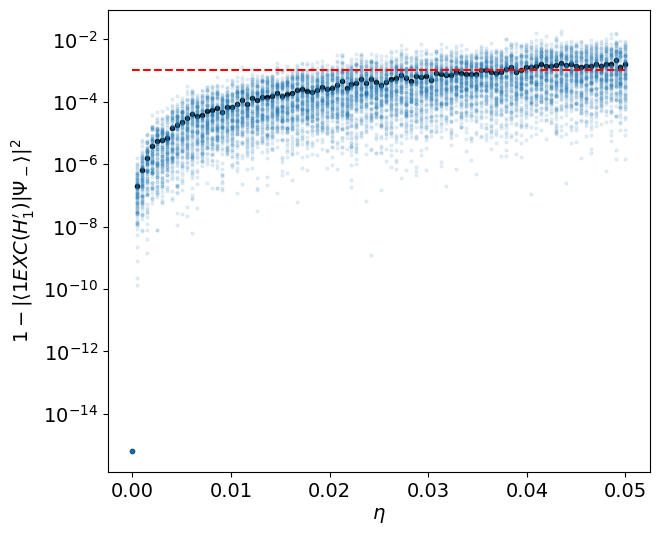}};
      \begin{scope}[x={(img.south east)}, y={(img.north west)}]
        \node[] at (0.34,0.9) {a)};
        \node[] at (0.88,0.3) {b)};
      \end{scope}
    \end{tikzpicture}
    \caption{Analysis on the impact of control imperfections. In a) we show the achieved infidelities considering 100 realisations of control errors for $J=1, t_0 = 20$ and $\varepsilon=0.02\cdot \Delta_{01}\simeq 0.001$. Each point corresponds to an individual noise realisation, from which then the mean and median are calculated. The red dashed line indicates the baseline performance of the perfectly homogeneous model. Plot b) shows the infidelity of the first excited state of the noisy $H_1^\prime$, $\ket{1EXC(H_1^\prime)}$, with the singlet state, $\ket{\Psi_-}$, as a function of $\eta$. The red dashed line corresponds to an infidelity of $10^{-3}$.}
    \label{fig:control_error_effect__deviation_1EXC_from_singlet}
\end{figure}

Fig.~\ref{fig:control_error_effect__deviation_1EXC_from_singlet}a shows the achieved errors in state preparation considering different noise strengths for $J=1, t_0=20$ and $\varepsilon = 0.02 \cdot \Delta_{01}$. Note that the $\Delta_{01}$ considered, both to determine the amplitude and the frequency of the oscillation, are the ones computed in the homogeneous case, since the deviations are unknown. The exact control precision will depend on the exact parameters of operation, but it can be estimated to be around $10^{-3}$ at most, as a ballpark value. We thus observe that it is possible to preserve a fidelity above 99.9\% in most cases even in the presence of systematic error as long as $\eta \lesssim 2\cdot 10^{-3}$.

To determine whether the controllability error can be mitigated, we study the deviation of the first excited state ($\ket{1EXC}$) of $H_1^\prime = \frac{1}{2}(H_{1a}^\prime + H_{1b}^\prime)$ as a function of $\eta$ in Fig.~\ref{fig:control_error_effect__deviation_1EXC_from_singlet}b. From this plot we learn that the fidelity with our target state still remains above 99.9\% for $\eta=0.01$, and thus we should be able to optimise our protocol for the particular noise realisation of the experiment, either by characterising the true $\Delta_{01}^\prime$ directly or by optimising the oscillation frequency to maximise the final fidelity. With this improvement, along with the potential optimisation of the adiabatic interpolation step beyond a linear ramp, our simulations suggest that preparation errors of $10^{-4}$ may be feasible with calibration and schedule optimisations.

\newpage
\section{Conclusions}

We have presented a novel protocol for the preparation of singlet states $\ket{\Psi_-}$ with an analogue device hosting the Ising model with transverse fields. This protocol exploits the fact that $\ket{\Psi_-}$ is the first excited state of a family of Hamiltonians that is naturally compatible with superconducting control primitives. 

The robust preparation of this state is achieved by means of a two-step protocol. The first part consists of an adiabatic interpolation preparing the ground state of an intermediate Hamiltonian with a large gap, facilitated by symmetry protection. The second step comprises a resonant population transfer between the ground and first excited states via a controlled oscillation of terms in the Hamiltonian which ensures that the relevant gap remains constant to first order.

For realistic parameters in the GHz regime, the protocol achieves fidelities comparable to standard, high-quality digital preparation while requiring similar total durations ($\sim 100$ ns). By reducing the oscillation amplitude and modestly increasing the protocol duration (by a factor of 2–4), infidelities in the $10^{-4}$ range become achievable in principle. Importantly, the protocol remains robust against moderate systematic control errors ($\eta \lesssim 10^{-3} - 10^{-2}$). Further optimisation, such as calibrating the actual excitation gap or employing non-linear adiabatic schedules, could enhance performance further. As an outlook, a realistic analysis of the impact of decoherence on this protocol is pending.

\bibliography{references}

@article{bennett_teleporting_1993,
	title = {Teleporting an unknown quantum state via dual classical and {Einstein}-{Podolsky}-{Rosen} channels},
	volume = {70},
	url = {https://link.aps.org/doi/10.1103/PhysRevLett.70.1895},
	doi = {10.1103/PhysRevLett.70.1895},
	number = {13},
	urldate = {2026-02-18},
	journal = {Physical Review Letters},
	author = {Bennett, Charles H. and Brassard, Gilles and Crépeau, Claude and Jozsa, Richard and Peres, Asher and Wootters, William K.},
	month = mar,
	year = {1993},
	note = {Publisher: American Physical Society},
	pages = {1895--1899},
}

@article{bennett_communication_1992,
	title = {Communication via one- and two-particle operators on {Einstein}-{Podolsky}-{Rosen} states},
	volume = {69},
	url = {https://link.aps.org/doi/10.1103/PhysRevLett.69.2881},
	doi = {10.1103/PhysRevLett.69.2881},
	number = {20},
	urldate = {2026-02-18},
	journal = {Physical Review Letters},
	author = {Bennett, Charles H. and Wiesner, Stephen J.},
	month = nov,
	year = {1992},
	note = {Publisher: American Physical Society},
	pages = {2881--2884},
}

@article{ekert_quantum_1991,
	title = {Quantum cryptography based on {Bell}'s theorem},
	volume = {67},
	url = {https://link.aps.org/doi/10.1103/PhysRevLett.67.661},
	doi = {10.1103/PhysRevLett.67.661},
	number = {6},
	urldate = {2026-02-18},
	journal = {Physical Review Letters},
	author = {Ekert, Artur K.},
	month = aug,
	year = {1991},
	note = {Publisher: American Physical Society},
	pages = {661--663},
}

@article{horodecki_quantum_2009,
	title = {Quantum entanglement},
	volume = {81},
	url = {https://link.aps.org/doi/10.1103/RevModPhys.81.865},
	doi = {10.1103/RevModPhys.81.865},
	number = {2},
	urldate = {2026-02-18},
	journal = {Reviews of Modern Physics},
	author = {Horodecki, Ryszard and Horodecki, Paweł and Horodecki, Michał and Horodecki, Karol},
	month = jun,
	year = {2009},
	note = {Publisher: American Physical Society},
	pages = {865--942},
}

@article{kwiat_new_1995,
	title = {New {High}-{Intensity} {Source} of {Polarization}-{Entangled} {Photon} {Pairs}},
	volume = {75},
	url = {https://link.aps.org/doi/10.1103/PhysRevLett.75.4337},
	doi = {10.1103/PhysRevLett.75.4337},
	number = {24},
	urldate = {2026-02-18},
	journal = {Physical Review Letters},
	author = {Kwiat, Paul G. and Mattle, Klaus and Weinfurter, Harald and Zeilinger, Anton and Sergienko, Alexander V. and Shih, Yanhua},
	month = dec,
	year = {1995},
	note = {Publisher: American Physical Society},
	pages = {4337--4341},
}

@article{king_quantum_2023,
	title = {Quantum critical dynamics in a 5,000-qubit programmable spin glass},
	volume = {617},
	copyright = {2023 The Author(s), under exclusive licence to Springer Nature Limited},
	issn = {1476-4687},
	url = {https://www.nature.com/articles/s41586-023-05867-2},
	doi = {10.1038/s41586-023-05867-2},
	number = {7959},
	urldate = {2026-02-18},
	journal = {Nature},
	author = {King, Andrew D. and Raymond, Jack and Lanting, Trevor and Harris, Richard and Zucca, Alex and Altomare, Fabio and Berkley, Andrew J. and Boothby, Kelly and Ejtemaee, Sara and Enderud, Colin and Hoskinson, Emile and Huang, Shuiyuan and Ladizinsky, Eric and MacDonald, Allison J. R. and Marsden, Gaelen and Molavi, Reza and Oh, Travis and Poulin-Lamarre, Gabriel and Reis, Mauricio and Rich, Chris and Sato, Yuki and Tsai, Nicholas and Volkmann, Mark and Whittaker, Jed D. and Yao, Jason and Sandvik, Anders W. and Amin, Mohammad H.},
	month = may,
	year = {2023},
	note = {Publisher: Nature Publishing Group},
	pages = {61--66},
}

@article{king_beyond-classical_2025,
	title = {Beyond-classical computation in quantum simulation},
	volume = {388},
	url = {https://www-science-org.sire.ub.edu/doi/10.1126/science.ado6285},
	doi = {10.1126/science.ado6285},
	number = {6743},
	urldate = {2026-02-18},
	journal = {Science},
	author = {King, Andrew D. and Nocera, Alberto and Rams, Marek M. and Dziarmaga, Jacek and Wiersema, Roeland and Bernoudy, William and Raymond, Jack and Kaushal, Nitin and Heinsdorf, Niclas and Harris, Richard and Boothby, Kelly and Altomare, Fabio and Asad, Mohsen and Berkley, Andrew J. and Boschnak, Martin and Chern, Kevin and Christiani, Holly and Cibere, Samantha and Connor, Jake and Dehn, Martin H. and Deshpande, Rahul and Ejtemaee, Sara and Farre, Pau and Hamer, Kelsey and Hoskinson, Emile and Huang, Shuiyuan and Johnson, Mark W. and Kortas, Samuel and Ladizinsky, Eric and Lanting, Trevor and Lai, Tony and Li, Ryan and MacDonald, Allison J. R. and Marsden, Gaelen and McGeoch, Catherine C. and Molavi, Reza and Oh, Travis and Neufeld, Richard and Norouzpour, Mana and Pasvolsky, Joel and Poitras, Patrick and Poulin-Lamarre, Gabriel and Prescott, Thomas and Reis, Mauricio and Rich, Chris and Samani, Mohammad and Sheldan, Benjamin and Smirnov, Anatoly and Sterpka, Edward and Trullas Clavera, Berta and Tsai, Nicholas and Volkmann, Mark and Whiticar, Alexander M. and Whittaker, Jed D. and Wilkinson, Warren and Yao, Jason and Yi, T. J. and Sandvik, Anders W. and Alvarez, Gonzalo and Melko, Roger G. and Carrasquilla, Juan and Franz, Marcel and Amin, Mohammad H.},
	month = apr,
	year = {2025},
	note = {Publisher: American Association for the Advancement of Science},
	pages = {199--204},
}

@article{hu_overcoming_2024,
	title = {Overcoming the coherence time barrier in quantum machine learning on temporal data},
	volume = {15},
	copyright = {2024 The Author(s)},
	issn = {2041-1723},
	url = {https://www.nature.com/articles/s41467-024-51162-7},
	doi = {10.1038/s41467-024-51162-7},
	number = {1},
	urldate = {2025-11-03},
	journal = {Nature Communications},
	author = {Hu, Fangjun and Khan, Saeed A. and Bronn, Nicholas T. and Angelatos, Gerasimos and Rowlands, Graham E. and Ribeill, Guilhem J. and Türeci, Hakan E.},
	month = aug,
	year = {2024},
	note = {Publisher: Nature Publishing Group},
	pages = {7491},
}

@article{karamlou_probing_2024,
	title = {Probing entanglement in a {2D} hard-core {Bose}–{Hubbard} lattice},
	volume = {629},
	copyright = {2024 The Author(s)},
	issn = {1476-4687},
	url = {https://www.nature.com/articles/s41586-024-07325-z},
	doi = {10.1038/s41586-024-07325-z},
	number = {8012},
	urldate = {2026-02-18},
	journal = {Nature},
	author = {Karamlou, Amir H. and Rosen, Ilan T. and Muschinske, Sarah E. and Barrett, Cora N. and Di Paolo, Agustin and Ding, Leon and Harrington, Patrick M. and Hays, Max and Das, Rabindra and Kim, David K. and Niedzielski, Bethany M. and Schuldt, Meghan and Serniak, Kyle and Schwartz, Mollie E. and Yoder, Jonilyn L. and Gustavsson, Simon and Yanay, Yariv and Grover, Jeffrey A. and Oliver, William D.},
	month = may,
	year = {2024},
	note = {Publisher: Nature Publishing Group},
	pages = {561--566},
}

@article{rosen_flat-band_2025,
	title = {Flat-{Band} ({De})localization {Emulated} with a {Superconducting} {Qubit} {Array}},
	volume = {15},
	url = {https://link.aps.org/doi/10.1103/PhysRevX.15.021091},
	doi = {10.1103/PhysRevX.15.021091},
	number = {2},
	urldate = {2026-02-18},
	journal = {Physical Review X},
	author = {Rosen, Ilan T. and Muschinske, Sarah and Barrett, Cora N. and Rower, David A. and Das, Rabindra and Kim, David K. and Niedzielski, Bethany M. and Schuldt, Meghan and Serniak, Kyle and Schwartz, Mollie E. and Yoder, Jonilyn L. and Grover, Jeffrey A. and Oliver, William D.},
	month = jun,
	year = {2025},
	note = {Publisher: American Physical Society},
	pages = {021091},
}

@article{xu_high-fidelity_2020,
	title = {High-{Fidelity}, {High}-{Scalability} {Two}-{Qubit} {Gate} {Scheme} for {Superconducting} {Qubits}},
	volume = {125},
	url = {https://link.aps.org/doi/10.1103/PhysRevLett.125.240503},
	doi = {10.1103/PhysRevLett.125.240503},
	number = {24},
	urldate = {2026-02-17},
	journal = {Physical Review Letters},
	author = {Xu, Yuan and Chu, Ji and Yuan, Jiahao and Qiu, Jiawei and Zhou, Yuxuan and Zhang, Libo and Tan, Xinsheng and Yu, Yang and Liu, Song and Li, Jian and Yan, Fei and Yu, Dapeng},
	month = dec,
	year = {2020},
	note = {Publisher: American Physical Society},
	pages = {240503},
}

@misc{an_zz-free_2025,
	title = {{ZZ}-{Free} {Two}-{Transmon} {CZ} {Gate} {Mediated} by a {Fluxonium} {Coupler}},
	url = {http://arxiv.org/abs/2511.02115},
	doi = {10.48550/arXiv.2511.02115},
	urldate = {2026-09-07},
	publisher = {arXiv},
	author = {An, Junyoung and Zhang, Helin and Ding, Qi and Ding, Leon and Sung, Youngkyu and Winik, Roni and Kim, Junghyun and Rosen, Ilan T. and Azar, Kate and Piñero, Renee DePencier and Gertler, Jeffrey M. and Gingras, Michael and Niedzielski, Bethany M. and Stickler, Hannah and Schwartz, Mollie E. and Wang, Joel I-j and Orlando, Terry P. and Gustavsson, Simon and Hays, Max and Grover, Jeffrey A. and Serniak, Kyle and Oliver, William D.},
	month = nov,
	year = {2025},
	note = {arXiv:2511.02115 [quant-ph]},
}

@article{li_realization_2024,
	title = {Realization of {High}-{Fidelity} {CZ} {Gate} {Based} on a {Double}-{Transmon} {Coupler}},
	volume = {14},
	url = {https://link.aps.org/doi/10.1103/PhysRevX.14.041050},
	doi = {10.1103/PhysRevX.14.041050},
	number = {4},
	urldate = {2026-02-17},
	journal = {Physical Review X},
	author = {Li, Rui and Kubo, Kentaro and Ho, Yinghao and Yan, Zhiguang and Nakamura, Yasunobu and Goto, Hayato},
	month = nov,
	year = {2024},
	note = {Publisher: American Physical Society},
	pages = {041050},
}

@article{wilk_entanglement_2010,
	title = {Entanglement of {Two} {Individual} {Neutral} {Atoms} {Using} {Rydberg} {Blockade}},
	volume = {104},
	url = {https://link.aps.org/doi/10.1103/PhysRevLett.104.010502},
	doi = {10.1103/PhysRevLett.104.010502},
	number = {1},
	journal = {Physical Review Letters},
	author = {Wilk, T. and Gaëtan, A. and Evellin, C. and Wolters, J. and Miroshnychenko, Y. and Grangier, P. and Browaeys, A.},
	month = jan,
	year = {2010},
	note = {},
	pages = {010502},
}

@book{sakurai_modern_2011,
	address = {Boston},
	edition = {2nd ed},
	title = {Modern quantum mechanics},
	isbn = {978-0-8053-8291-4},
	publisher = {Addison-Wesley},
	author = {Sakurai, J. J. and Napolitano, Jim},
	year = {2011},
}

\end{document}